# Hallmarks of Human-Machine Collaboration
A framework for assessment in the DARPA Communicating with Computers program

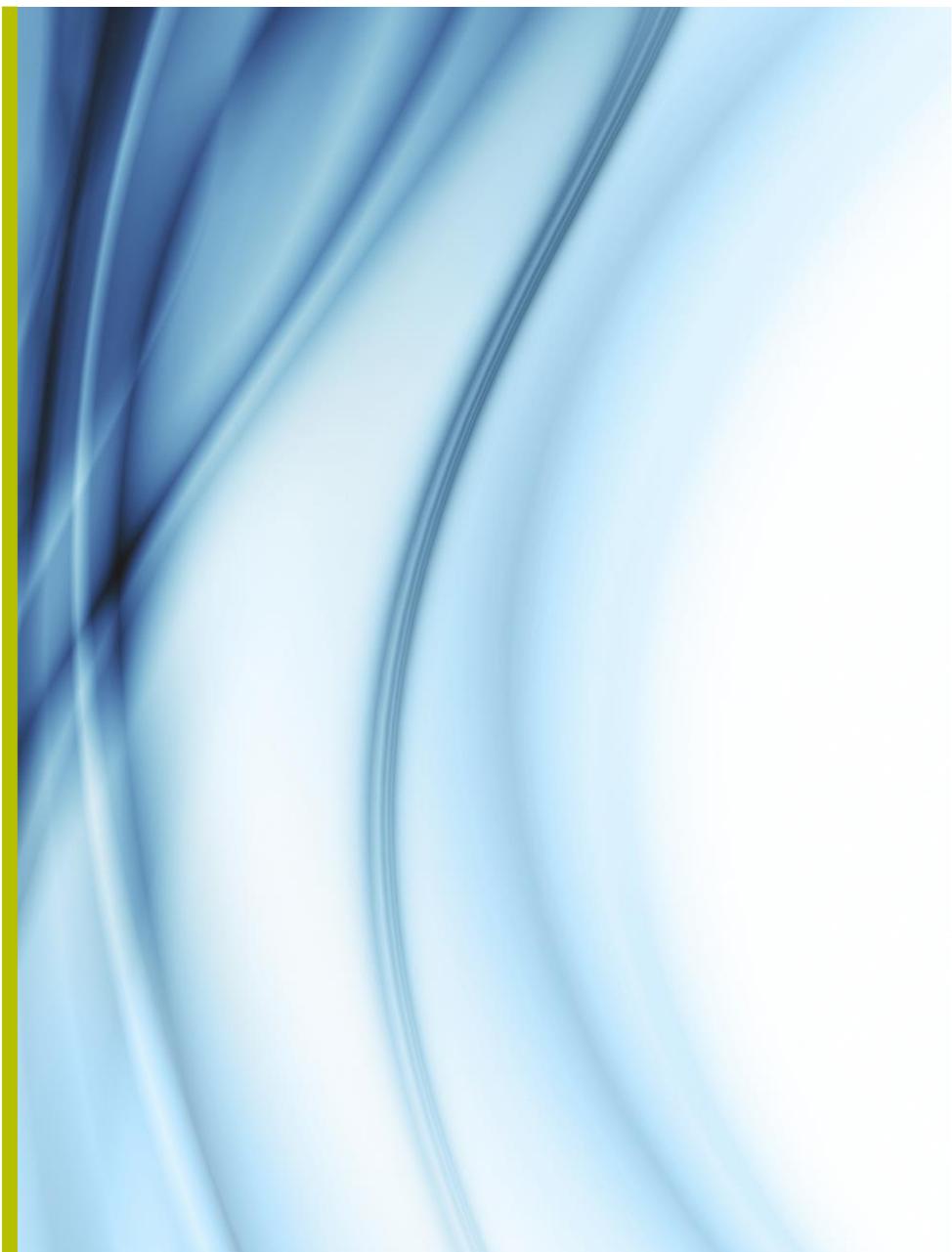

MITRE

This page intentionally left blank.

# Introduction

There is a growing desire to create computer systems that can communicate effectively to collaborate with humans on complex, open-ended activities, often with a particular goal. Assessing this type of system presents significant challenges. This report describes a framework for evaluating collaborative computer systems engaged in open-ended complex scenarios where evaluators do not have the luxury of comparing performance to a single right answer. This framework was developed in the context of the DARPA Communicating with Computers (CwC) program, and has been used in the evaluation of communication in disparate human-machine creative collaborations, including story and music generation, interactive block building, and exploration of molecular mechanisms in cancer.

These activities are fundamentally different from the more constrained tasks performed by most contemporary personal assistants and other communicative systems such as information retrieval, scheduling, setting alarms, or turning on lights. Creative collaboration activities are generally open-ended, with no single correct solution, and in many cases no obvious completion criteria. They explore human-machine partnership and creativity via (possibly multi-modal) conversation. These collaborations frequently involve building structures or compositions through multiple steps, and consequently dialogues for them require the use of shared context about what is being built and what has been discussed before. Success in these endeavors relies heavily on world knowledge and common sense. Finally, the complex nature of these activities can make system robustness particularly challenging. Ideally, the resulting creative collaborative systems would enable humans to communicate with machine collaborators easily, enjoyably, and efficiently, and together generate high-quality creations, products and/or outcomes.

In our role as the test and evaluation (T&E) team for CwC, we focused on the major program goals to identify the Key Properties that must be exhibited by systems to be considered successful. From there we identified "Hallmarks" of success – capabilities and features that evaluators can observe that would be indicative of progress toward achieving a Key Property. In addition to being a framework for assessment, the Key Properties and Hallmarks are intended to serve as goals in guiding research direction.

This report describes the Hallmarks of Human-Machine Collaboration developed for the DARPA Communicating with Computers program.

# Goals of the DARPA CwC Program

The goals of the CwC Program are summarized by the following description from DARPA's website:

> *The Communicating with Computers (CwC) program aims to enable symmetric communication between people and computers in which machines are not merely receivers of instructions but collaborators, able to harness a full range of natural modes including language, gesture and facial or other expressions. For the purposes of the CwC program, communication is understood to be the sharing of complex ideas in collaborative contexts. Complex ideas are assumed to be built from a relatively small* *set of elementary ideas, and language is thought to specify such complex ideas—but not completely, because language is ambiguous and depends in part on context, which can augment language and improve the specification of complex ideas. Thus, the CwC program will focus on developing technology for assembling complex ideas from elementary ones given language and context.*[1]

## From Goals to Key Properties and Hallmarks

From the goals of the CwC program we identified eight Key Properties of successful CwC systems. These are the characteristics that DARPA and MITRE identified as desirable and/or necessary for the types of systems whose development the CwC program wished to support. However, the Key Properties are multi-faceted, so in order to assess how well systems embody these properties, we considered what observable features a system succeeding in these dimensions would exhibit. These observables are our assessment Hallmarks. For convenience, we have numbered the Hallmarks within each Key Property using a two-letter abbreviation of the Key Property name, followed by a number.

For example, for the CwC program, there was a program goal of enabling bidirectional communication "between humans and computers in which machines are not merely receivers of instructions but collaborators." Systems achieving this goal would have the Key Property we named "**Mutual Contribution of Meaningful Content**," one aspect of which is the *machine's knowledge of when to act and how much to contribute*. The observable Hallmarks we look for to determine how well that is being achieved include:

- MC-1. Partners each take multiple turns in the interaction
- MC-2. Each partner knows when to communicate and/or take actions
- MC-3. Machine responses are of an appropriate length and level of detail
- MC-4. The machine takes initiative when appropriate
- MC-5. If the human grants autonomy, the machine responds appropriately

Hallmark satisfaction and progress can be assessed in a number of ways. To assess Hallmarks like MC-1 through MC-5 above, we perform a detailed analysis of system interaction logs. This is typically not an all-or-nothing assessment, but rather a quantification of how frequently each assessed Hallmark is achieved or missed in a given set of interactions.

Other Hallmarks require us to survey the human partners. For instance, one of the Hallmarks under the CwC Key Property of "**Consistent Human Engagement**" is:

- HE-1. Human partners can communicate successfully in a way that is comfortable

This is something we assess via the Likert scale survey item:

> I was able to communicate successfully in a way that was comfortable

---

[1] https://www.darpa.mil/program/communicating-with-computers

We might also survey third parties (often via crowdsource platforms such as Amazon Mechanical Turk) to observe whether or to what extent Hallmarks like this one are achieved:

> SC-4. Doing the task together results in a more interesting, creative, or otherwise better product

## Key Properties and Hallmarks for CwC

Here we present the eight Key Properties that exemplary CwC systems should embody, along with the observable Hallmarks of success, divided into sub-categories. The first Key Property, **Successful Collaboration**, is a higher-level Property than the others. The remaining seven Key Properties represent facets of successful CwC systems that contribute to the system's ability to satisfy the **Successful Collaboration** Property. Note that the examples given below are not intended to convey the current capability of a system but to illustrate exemplars or violations of Hallmarks discovered along the way during the development process. Some examples come from real CwC systems, whereas others are notional.

In the sections below we present the Key Properties and Hallmarks for CwC. Each Key Property has a high-level description (in blue) followed by a set of italicized sub-categories of the Property which are used to organize the set of Hallmarks for that Property. The Key Properties and their subcategories do have some overlap (especially with Successful Collaboration, which builds from all the lower-level Properties). For Hallmarks that may support more than one Key Property, we have used our judgment to select the best fit.

## Successful Collaboration

Satisfying creative collaborations can take place in which machines are not merely receivers of instructions but are full collaborators

### *Efficient, collaborative project completion*

> SC-1. The human-machine team completes projects
> SC-2. Task completion is efficient

### *Worthwhile collaboration*[2]

> SC-3. It's easier to do the activity together than alone
> SC-4. Doing the activity together results in a more interesting, creative, or otherwise better product
> SC-5. It is more enjoyable to do the activity together than alone

### *Human satisfaction*

> SC-6. Overall, the human is satisfied with the collaboration
> SC-7. Humans want to interact further with the machine

---
[2] To achieve **Worthwhile collaboration,** a system should demonstrate at least one of the three Hallmarks in this subcategory; typically we do not see or expect to see all three for any given system.

# Robustness

## Efficient task-based interaction proceeds smoothly as long as the human wants to, without resets

### *Software reliability and consistency*

RO-1. The interaction proceeds without the need for resets (no crashes/hangs)
RO-2. All expected capabilities are online and working as expected
RO-3. The machine produces consistent content
RO-4. The machine responds with appropriate actions

### *Ability of human and machine to understand diverse communications*

RO-5. The human's communication is correctly interpreted by the machine
RO-6. The machine handles multiple phrasings or forms for similar requests
RO-7. The machine's communication is readily interpreted by the human
RO-8. The machine communicates effectively via multiple modalities (e.g., speech, gesture, text, audio, facial expressions), as appropriate
RO-9. The machine correctly interprets multiple communication modalities
RO-10. The set of inputs the machine can interpret is enough to support the full functionality of the system

*The machine rearranges a score in response to human's manipulation of a note (click and drag):*

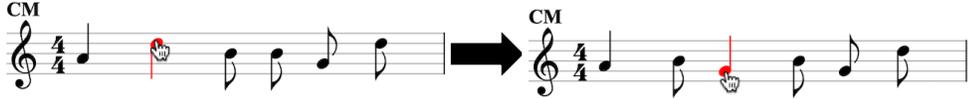

*The machine generates a score in response to the human's natural language request:*

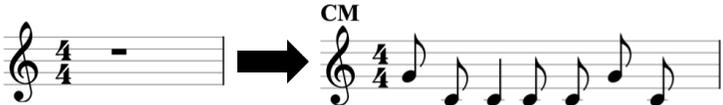

**Human:** *Generate 1 measure.*

**Machine:** *Operation Successful. I think you wanted to: generate 1 measure.*

**Figure 1. Multi-modal interaction: Direct manipulation versus verbal request (RO-8)[3]**

---

[3] Figure content generated by the MUSICA system [1-2].

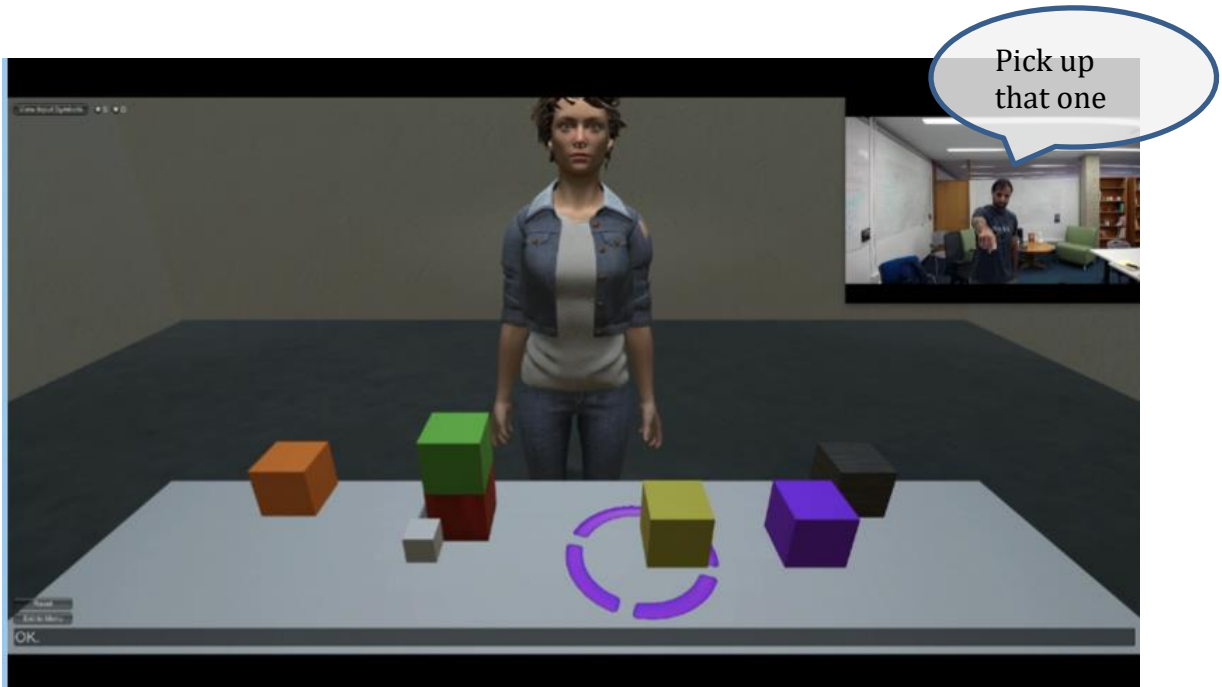

**Figure 2. Multi-modal interaction: Speech with (pointing) gesture (RO-9).**[4] **The human is pointing from his perspective at the place on the virtual table indicated by the purple circle. In addition to aligning the gesture with the corresponding spoken command, the machine must also correctly interpret the deictic (in this case, spatial) context of the gesture (CA-6).**

### *Ability of the machine to move the conversation forward past misunderstandings*

RO-11.  The machine copes with errors in the human's input
RO-12.  The machine repairs interactions and/or supports human repair
RO-13.  The machine provides helpful/actionable error messages
RO-14.  The machine asks clarifying questions as needed
RO-15.  The machine responds appropriately to humans' requests for clarification

> **Human:** *Is KRAS a transcriptoin factor?*
>
> **Machine:** *I have no evidence for KRAS being a transcription factor.*

**Figure 3. Correction of the human's spelling errors (RO-11)**[5]

---

[4] Figure content generated by the Diana system [3-5]
[5] Figure content generated by the Bob with Bioagents system [6]

> *Sorry, I didn't understand what you wanted to do. Try typing 'help' for a list of commands.*
>
> *It looks like you are trying to transpose something. I need to know what to transpose and by how much.*
>
> *Sorry, I didn't catch that. Can you rephrase please?*
>
> *I am having trouble, possibly because I don't know what 'those' refers to.*
>
> *That structure isn't symmetric. Try moving a block to make it symmetric.*

**Figure 4. Helpful/actionable error messages (RO-13)[6]**

> **Human command:** *The half note starting on beat 3 of bar 2 should be one octave up.*
>
> **Machine response:** *line 1:14 mismatched input 'start' expecting '***'*
> *Operation Failure. Command given: the half note starting on beat 3 of bar 2 should be one octave up. It looks like you are trying to transpose something. I need to know what to transpose and by how much. If you need more information on command format, type 'help transpose'.*

**Figure 5. Clarification exchange (RO-14)[7]**

> **Human**: *What regulates MEK?*
>
> **Machine**: *By MEK, do you mean MAP2K1 or MAP2K2?*
>
> **Human:** *MAP2K1*

**Figure 6. Clarifying question and answer (RO-14)[8]**

## Mutual Contribution of Meaningful Content

Each participant makes meaningful contributions to the session, and either participant can take or cede initiative

### *Machine's knowledge of when to act and how much to contribute*

    MC-1.    Partners each take multiple turns in the interaction

---

[6] Figure content generated by the MUSICA system [1-2] (1st and 2nd sentences), the Bob with Bioagents system [6] (3rd and 4th sentences), and SMILEE [7] (5th sentence)
[7] Figure content generated by the MUSICA system
[8] Figure content generated by the Bob with Bioagents system

MC-2. Each partner knows when to communicate and/or take actions
MC-3. Machine responses are of an appropriate length and level of detail
MC-4. The machine takes initiative when appropriate
MC-5. If the human grants autonomy, the machine responds appropriately

> **Machine:** *It looks like a required activity may be missing, say "Active AGT activates ERK." to add it*

Figure 7. Machine takes intitative (MC-4)[9]

> **Machine:** *What color block should I use?*
>
> **Human:** *You choose*
>
> **Machine:** *Okay I will use a red block.*

Figure 8. Exchange in which human grants autonomy to machine (MC-5)[10]

### *Appropriate and collaborative contributions*

MC-6. The machine makes meaningful contributions to the interaction
MC-7. The machine enables the human to make meaningful contributions to the interaction
MC-8. Partners negotiate or collaboratively shape goals or approaches

## Consistent Human Engagement

Humans find engaging with machine comfortable, useful, fun, inspiring, and/or rewarding

### *Comfortable Interaction*

HE-1. Human partners can communicate successfully in a way that is comfortable
HE-2. The human is satisfied with the pacing/tempo of the interaction
HE-3. The human is satisfied with how things progress during the interaction
HE-4. The machine communicates without creating undue distraction

---

[9] Figure content generated by the Bob with Bioagents system [6]
[10] Figure content provided by Paul Cohen

> **Human:** *Jim slouched in the corner, feeling sorry for the patient in front of him.*
>
> **Machine:** *"This is ridiculous," said Duke.*
>
> **Human:** *"Yesterday I felt fine, and now you're telling me I'm at death's door?!"*
>
> **Machine:** *"We'll take care of Furble tomorrow,*
>
> **Human:** *" the doctor said. "You've named my tumor?! Duke shrieked.*
>
> **Machine:** Yeah,
>
> **Human:** *" replied the doctor coolly. "We've found that anthropomorphizing tumors helps people in your position come to terms with their condition more easily."*

**Figure 9. Satisfactory progression of the interaction *and* The machine produces content that is interesting, novel, useful, and/or creative (HE-3, HE-5). This story was written in response to an image prompt that depicts a doctor talking to patient in his offce as a man stands against a wall behind him. In interactive story generation, humans may be able to incorporate machine suggestions that appear random or unrelated to the previous context through imaginative means that yield a more creative story than would result from more obvious or expected machine contributions.[11]**

> *Ok. It should be true that the MIDDLE column is MIDDLE*
>
> *Hmm, from what I understood, it should be true that the every columns's height is equal to the all of the the every columns and the every columns's height is equal to the all of the the every columns. However, that doesn't match an example you showed me. Do you still want to add this constraint?*

**Figure 10. Distracting machine communications (negative examples of HE-4)[12]**

---

[11] Figure content generated by the University of Washington interactive story generation system [8]
[12] Figure content generated by an IHMC prototype system for structure learning in blocks world [9]

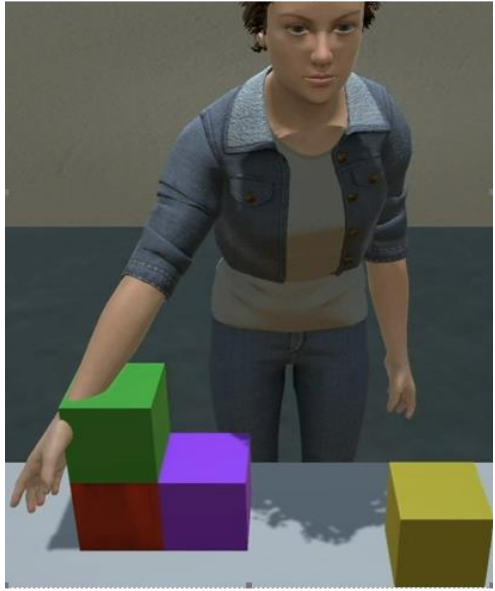

**Figure 11. Distracting (physically impossible) machine-generated visualization (negative example of HE-4)**[13]

### *Machine's ability to evoke and Inspire*

- HE-5. The machine produces content that is interesting, novel, useful, and/or creative
- HE-6. Machine inspires new ideas in the human partner
- HE-7. The machine evokes emotional responses (other than frustration)

## Context-awareness

Both partners can communicate efficiently by referencing and understanding contexts, including the linguistic, conversational, and deictic context; task context; goal context; self knowledge; the partner's abilities; and world/domain knowledge

### *Linguistic and/or deictic context-awareness*

- CA-1. The machine recognizes co-referring mentions of previously mentioned entities (co-reference resolution)
- CA-2. The machine recognizes references to previously created units or sub-parts
- CA-3. The machine generates references to previously created units or sub-parts
- CA-4. The machine correctly interprets a term defined by a human partner
- CA-5. The machine uses a term defined by a human partner
- CA-6. The machine correctly interprets and correctly uses deictic references (i.e., references situated in time and/or place, such as by pointing)

---

[13] Figure content generated by the Diana system [3-5]

> **Machine:** *I found that STAT3 regulates the following proteins:...*
>
> **Human:** *Does PLX-4720 target any of those?*
>
> **Machine:** *Of those 421 proteins, I found six targets of PLX-4720:....*

**Figure 12. Recognition of co-referring mentions of previously mentioned entities (CA-1)[14]**

> **Human:** "*I think we need some giant cats," Paul said to his wife Joan.*
>
> **Machine:** "*I don't understand," Joan said. "I don't even know why we're calling in giant cats."*

**Figure 13. Generation of co-referring mentions of previously mentioned entities (CA-3)[15]**

### *Pragmatic context-awareness*

CA-7.   The machine indicates when a requested action is outside of its capabilities
CA-8.   The machine communicates its situationally relevant capabilities to the human
CA-9.   The machine indicates that it doesn't understand what a particular entity/action/word/gesture is when appropriate
CA-10.  The machine maps its interpretation to its appropriate capabilities
CA-11.  The machine tailors responses to the human partner's apparent capabilities
CA-12.  The machine tailors responses to the human partner's stated or implied goals
CA-13.  Human partners tailor responses to machine's stated or implied goals

> **Human:** *I want to find out how Jak1 increases the amount of IL2*
>
> **Machine:** *Okay, from now on I will assume that the goal of our model is to explain JAK1 increases the amount of IL2.*
>
> **Human:** *Active JAK1 activates STAT3*
>
> **Machine:** *Okay, I added Active JAK1 activates STAT3 to the model.*
>
> **Human:** *Active STAT increases the amount of IL2*
>
> **Machine:** *Okay, I added active STAT3 increase sthe amount of IL2 to the model. Hooray! Our model can now explain our goal through these mechanisms: Active JAK1 activates STAT3. Active STAT3 increases the amount of IL2.*

**Figure 14. Machine tailors its response to a goal stated by the human partner (CA-12)[16]**

---

[14] Figure content generated by the Bob with Bioagents system [6]
[15] Figure content generated by the University of Washington interactive story generation system [8]
[16] Figure content generated by the CLARE system [10]

*Situational context-awareness*

- CA-14. The machine responds appropriately to human references and actions in the context of the evolving situation (includes anything built and pieces available)
- CA-15. The machine's contributions to the interaction are consistent, relevant, and build from turn to turn (i.e. are coherent)

> **Human**: *What pathways are shared by IGTB1 and AGT?*
>
> **Machine:** *I am having trouble because I don't know what 'IGTB1' is.*

**Figure 15. Machine indicates it does not understand a particular word (CA-9)**[17]

> *Pour* ==batter== *into prepared pan.*
>
> *Bake in the preheated oven for 60 minutes, or until a toothpick inserted into the center of the* ==cake== *comes out clean.*

**Figure 16. References that reflect the evolving situation (CA-14)**[18]

*Appropriate use of world/domain knowledge*

- CA-16. The machine applies world/domain knowledge as required
- CA-17. Machine contributions are generally consistent with common sense
- CA-18. Machine demonstrates understanding of a broad range of vocabulary

> **Human recipe title:** *marble chocolate chip* ==cupcakes==
>
> **Machine recipe instruction:** *Spoon batter into a greased* ==8 " square baking pan==.

**Figure 17. Violation of cooking domain knowledge (negative example of CA-16)**[19]

# Provision of Rationale

## The machine can expose its reasoning, sources, and methods

---

[17] Figure content generated by the Bob with Bioagents [6]
[18] Figure content generated by the University of Washington recipe generation system [11-13]
[19] Figure content generated by the University of Washington recipe generation system

*Logging*

RA-1. The machine can provide a human-interpretable output (trace), showing processes implemented by the computer sufficient for evaluators' use

*Ability to explain rationale*

RA-2. The machine provides information about the sources and methods it used to generate a response or action
RA-3. The machine provides reasons for its responses and actions that make sense to the human partner
RA-4. The machine answers questions about its reasoning, sources, and/or methods

*Human's trust is appropriately calibrated*

RA-5. The human partner knows when to trust the machine's responses and actions

## Habitability

Humans easily learn to use language and/or gestures that the machine can correctly interpret and act upon

*Shaping*

HA-1. The machine offers information that helps human partners produce utterances and/or gestures that the machine understands and are consistent with its capabilities
HA-2. The machine offers information that helps human partners produce utterances and/or gestures that are consistent with the partners' goals and move the collaborative effort forward
HA-3. The machine models appropriate use by generating utterances and/or gestures that it can interpret
HA-4. The machine answers questions about its capabilities
HA-5. The machine uses and understands language consistently (across capabilities)

> **Human:** *Does Niclosamide change the amount of PI3K in the model?*
>
> **Machine:** *Your question is similar to "Does ELK1 decrease the amount of FOS in the model?". Try asking it that way.*
>
> **Human:** *Does Niclosamide decrease the amount of PI3K in the model?*
>
> **Machine:** *Yes, PI3K decreased in my simulation.*

**Figure 18. Machine recognizes that a question is about a capability it has and suggests phrasing that it can understand (HA-1)[20]**

---

[20] Figure content generated by the CLARE system [10]

*Learnability*

    HA-6.    Human partners are able to quickly learn to use the system effectively

    HA-7.    The collaborative effort moves forward with minimal need for repeats or digressions

    HA-8.    The number of repair sub-dialogues decreases over time for each human partner

## Appropriate Use of Elementary Composable Ideas (ECIs)

Uses and composes elements of a shared set of elementary ideas to represent more complex concepts

*Program-wide sharing*

**Program-level Goal**

    EC-1.    Program produces a set of ECIs used and shared by multiple systems and use cases.

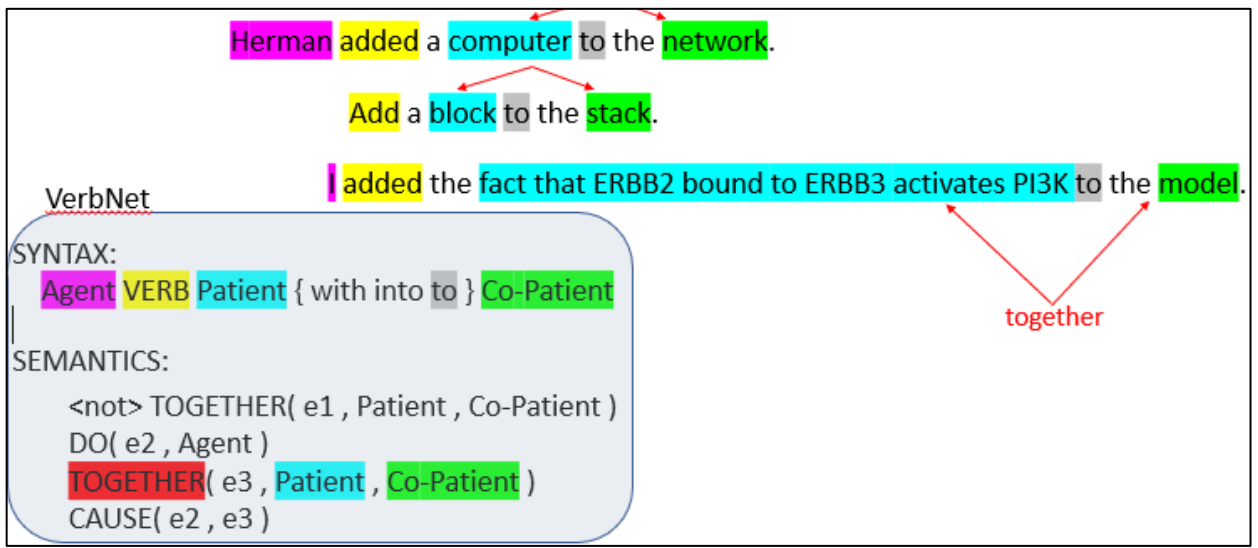

**Figure 19.** VerbNet enables consistent representation of events (such as adding something to something else) across domains and use cases (EC-1)[21]

---

[21] VerbNet (VN) is an on-line network of English verbs that links their syntactic and semantic patterns. It is available at https://verbs.colorado.edu/verbnet/ [14-16]

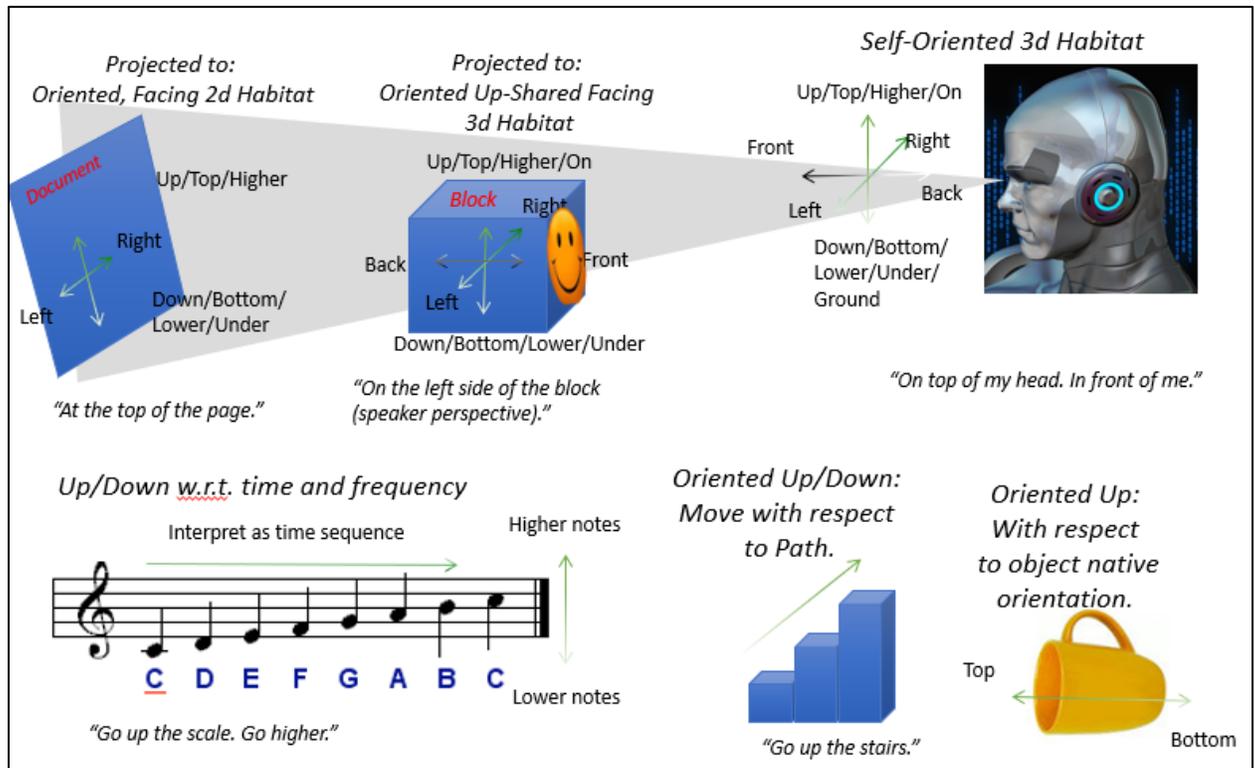

**Figure 20. The machine interprets the meaning of "up" based on context (EC-3)[22]**

## *Representation*

EC-2. The machine uses ECIs to represent concepts
EC-3. The machine uses non-domain-specific ECIs and grounds them according to context.

## *Composition*

EC-4. Humans teach a new concept by presenting it as a composition of known elementary ideas (ECIs)
EC-5. The machine can learn (or infer) the meaning of a new word or concept without explicit human instruction
EC-6. The machine introduces a new concept by presenting it as a composition of known elementary ideas (ECIs)

---

[22] Figure content provided by the Communicating with Computers Program ECI Working Group [17]

> **Human:** *I want to teach you a corner* <Human begins teaching a basic structure>
> **Machine**: *Can you tell me something about the structure?*
> **Human**: *The width is two*
> **Machine:** *Ok. It should be true that width is equal to 2*
> **Machine:** *Is this a correct example?* <System shows example>
> **Human**: *No. The second row has one block*
> **Machine:** *Ok. It should be true that the second row's number is equal to 1*
> **Machine:** *Can you tell me anything else about it?*
> **Human:** *We're done.*
> **Machine:** *Okay, thank you for teaching me this. If you need to do anything else, let me know.*

**Figure 21. The human teaches a new concept by presenting it as a composition of known elementary ideas (ECI-4)**[23]

# Conclusions

The Hallmarks approach offers a flexible approach to the assessment of complex computer systems when the notion of checking for "correct" outputs does not make sense. Hallmarks can be used as objectives in the development cycle, and in evaluation in user studies. The Hallmarks are similar to heuristics used elsewhere, but this set differs from others because of its focus on creative collaborative human-machine interactions. The CwC Hallmarks are defined with a level of generality that permit them to be applied to systems with very diverse purposes, from narrative, musical, and video composition to construction with blocks to biological model simulation. A subsequent paper will present how these hallmarks were used in evaluation in the CwC program.

The Hallmarks approach to evaluation has also been employed on the DARPA World Modelers program, which has very different types of goals than the CwC program. This demonstrates the flexibility of this type of framework for evaluation.

We have found that the exercise of being explicit about goals can help focus research, and has provided a structure for organizing presentations about system accomplishments. This approach to assessment appears to be an effective means of steering research in the direction of achieving program goals and objectives, and its flexibility allows it to grow and develop with evolving program goals.

# Acknowledgements

We thank Paul Cohen, the original DARPA CwC program manager, for thoughtful discussion and inputs to the initial set of CwC Key Properties and Hallmarks. We thank subsequent program managers Dave Gunning and Bruce Draper, and the program SETA Bill Bartko for continued helpful discussion and suggestions on the program Hallmarks as the program evolved. We thank the CwC Performer teams for insightful feedback into the Hallmarks and Hallmark-based evaluations throughout the duration of the program, many of whom have also provided illustrative examples for this document. Thanks to performer team members: James Allen, Mohamed Amer, Funda Durupinar Babur, John Bachman, Ross Beveridge, Yonatan Bisk, Rusty Bobrow, Antoine Bosselut, Susan Brown, Mark Burstein, Jan Buys,

---

[23] Figure content from an IHMC system for structure learning in blocks world [9]

Yejin Choi, Elizabeth Clark, Brent Cochran, Emek Demir, Chris Donahue, Lucian Galescu, Marjan Ghazvininejad, Jonathan Gordon, Ben Gyori, He He, Jerry Hobbs, Julia Hockenmaier, Ari Holtzman, Prashant Jayannavar, Ben Kane, Gene Kim, Kevin Knight, Nikhil Krishnaswamy, Percy Liang, Lara Martin, David McDonald, Tim Meo, Nasrin Mostafazadeh, Anjali Narayan-Chen, Sriraam Natarajan, Martha Palmer, Nanyun (Violet) Peng, Ian Perera, Georgiy Platonov, James Pustejovsky, Donya Quick, Hannah Rashkin, Mark Riedl, Dan Roth, Jaime Ruiz, Maartin Sap, Len Schubert, Noah Smith, Peter Sorger, Amir Tamrakar, Choh Man Teng, Kelland Thomas, Ralph Weischedel, Rowan Zellers, as well as others whose contributions may have been more in the background, but who nonetheless contributed to the successes of their teams.